\documentclass[floats,floatfix,showpacs,amssymb,prd,twocolumn,superscriptaddress,nofootinbib,reprint,preprintnumbers]{revtex4-2}

\usepackage{amssymb,amsmath,verbatim,mathtools,needspace,enumitem,etoolbox,graphicx,physics,afterpage,xspace,tabularx,lmodern,multirow}

\usepackage[caption=false]{subfig}   
\usepackage{ragged2e}                
\usepackage[babel=true]{microtype}
\usepackage{booktabs}
\usepackage{siunitx}
\usepackage{gensymb}
\usepackage{ulem}  
\usepackage[dvipsnames, usenames]{xcolor}
\definecolor{linkcolor}{rgb}{0.0,0.3,0.5}
\usepackage[unicode, colorlinks=true, linkcolor=linkcolor, citecolor=linkcolor, filecolor=linkcolor, urlcolor=linkcolor, linktocpage, breaklinks]{hyperref}
\usepackage[all]{hypcap}
\usepackage[T1]{fontenc}
\usepackage[utf8]{inputenc}
\usepackage[usenames,dvipsnames]{xcolor}

\hypersetup{colorlinks=true,citecolor=romared,linkcolor=romared,urlcolor=romared}

\usepackage{multirow}
\usepackage{appendix}
\usepackage{xcolor}
\usepackage{feynmp-auto}
\definecolor{romared}{RGB}{142,0,28}
\definecolor{tabblue}{RGB}{31, 119, 180}
\definecolor{darkblue}{RGB}{0, 0, 120}
\definecolor{tabred}{RGB}{214, 39, 40}
\definecolor{tabgreen}{RGB}{44, 160, 44}
\definecolor{tabgray}{RGB}{100, 100, 100}
\definecolor{goldenrod}{RGB}{218, 165, 32}
\definecolor{taborange}{RGB}{255, 127, 14}
\definecolor{tabbrown}{RGB}{128, 0, 0}
\definecolor{tabpink}{RGB}{255, 141, 161}
\definecolor{tabpurple}{RGB}{148, 103, 189}
\DeclareSIUnit{\parsec}{pc}
\DeclareSIUnit{\Mpc}{\mega\parsec}

\DeclareRobustCommand{\SkipTocEntry}[4]{}

\usepackage{mathtools}
\usepackage{float}
\usepackage{placeins}
\usepackage{aas_macros}
\usepackage{makecell}
\usepackage{soul}

\usepackage{lipsum}
\usepackage{orcidlink}

\begin{document}

\preprint{\hbox{MIT-CTP/5923, UTWI-29-2025, NORDITA-2025-049}}

\title{How Theory-Informed Priors Affect DESI Evidence for Evolving Dark Energy}

\author{Michael W. Toomey\,\orcidlink{0000-0003-1205-4033}}
\email{mtoomey@mit.edu}
\affiliation{Center for Theoretical Physics -- a Leinweber Institute, Massachusetts Institute of Technology, Cambridge, MA 02139, USA}

\author{Gabriele Montefalcone\,\orcidlink{0000-0002-6794-9064}}
\email{montefalcone@utexas.edu}
\affiliation{Texas Center for Cosmology and Astroparticle Physics, Weinberg Institute for Theoretical Physics, Department of Physics, The University of Texas at Austin, Austin, TX 78712, USA}

\author{Evan McDonough\,\orcidlink{0000-0002-2130-3903}}
\email{e.mcdonough@uwinnipeg.ca}
\affiliation{Department of Physics, University of Winnipeg, Winnipeg MB, R3B 2E9, Canada}

\author{Katherine Freese\,\orcidlink{0000-0001-9490-020X}}
\email{ktfreese@utexas.edu}
\affiliation{Texas Center for Cosmology and Astroparticle Physics, Weinberg Institute for Theoretical Physics, Department of Physics, The University of Texas at Austin, Austin, TX 78712, USA}
\affiliation{The Oskar Klein Centre, Department of Physics, Stockholm University, AlbaNova, SE-10691 Stockholm, Sweden}
\affiliation{Nordic Institute for Theoretical Physics (NORDITA), 106 91 Stockholm, Sweden}

\begin{abstract}
\noindent Recent measurements of baryon acoustic oscillations (BAO) from the Dark Energy Spectroscopic Instrument (DESI) have been interpreted to suggest that dark energy may be evolving. 
In this work, we examine how prior choices affect such conclusions. Specifically, we study the biases introduced by the customary use of uniform priors on the Chevallier--Polarski--Linder (CPL) parameters, $w_0$ and $w_a$, when assessing evidence for evolving dark energy. To do so, we construct theory-informed priors on $(w_0, w_a)$ using a normalizing flow (NF), trained on two representative quintessence models, which learns the distribution of these parameters conditional on the underlying $\Lambda$CDM parameters. In the combined {\it Planck} CMB + DESI BAO  analysis we find that the apparent tension with a  cosmological constant in the CPL framework can be reduced from $\sim 3.1\sigma$ to $\sim 1.3\sigma$ once theory-informed priors are applied, rendering the result effectively consistent with $\Lambda$CDM. For completeness, we also analyze combinations that include Type Ia supernova data, showing similar shifts toward the $\Lambda$CDM limit.  Taken together, the observed sensitivity to prior choices in these analyses arises because uniform priors -- often mischaracterized as ``uninformative'' -- can actually bias inferences toward unphysical parameter regions. Consequently, our results underscore the importance of adopting physically motivated priors to ensure robust cosmological inferences, especially when evaluating new hypotheses with only marginal statistical support. Lastly, our NF-based framework achieves these results by post-processing existing MCMC chains, requiring $\approx 1$~hour of additional CPU compute time on top of the base analysis -- a dramatic speedup over direct model sampling that highlights the scalability of this approach for testing diverse theoretical models.
\end{abstract}

\maketitle

\section{Introduction}

\noindent Recent measurements of baryon acoustic oscillations (BAO) from the Dark Energy Spectroscopic Instrument (DESI)  from their second data release (DR2), combined with datasets such as cosmic microwave background (CMB) anisotropies from \textit{Planck} and Type Ia supernovae, have been interpreted as  potential deviations from the standard $\Lambda$CDM cosmological model~\cite{DESI:2024mwx,DESI:2025zgx}. In particular,  when analyzed using the Chevallier-Polarski-Linder (CPL) parameterization for the DE equation of state~\cite{Chevallier:2000qy,Linder:2002et}:
\begin{equation}\label{eq:cpl}
    w_{\rm CPL}(z) = w_0 + w_a \frac{z}{1+z}
\end{equation}
and combining DESI DR2 BAO data with CMB observations, there is moderate evidence favoring a dynamical dark energy (DE) over a cosmological constant, with a significance exceeding $3\sigma$.

 This preference is further strengthened when incorporating supernovae datasets.\footnote{ Though a recent full shape analysis suggests this evidence is weakened~\cite{Chudaykin:2025aux}.} Notably, the parameter space driving this preference points to a phantom crossing, where the effective equation of state starts with $w < -1$ at early times and transitions to $w > -1$ before reaching the present epoch ($z = 0$)~\cite{DESI:2025fii}.  
 
Following the DESI results, numerous studies have reexamined the evidence for time-varying DE using alternative approaches to the standard CPL parameterization, finding that while the reconstructed evolution of $w(z)$ appears broadly consistent across methods, the statistical significance of deviations from $\Lambda$ varies substantially~(e.g.~\cite{DESI:2024kob,Luongo:2024fww,Shajib:2025tpd,Jiang:2024xnu,DESI:2025fii,Shlivko:2025fgv,Efstathiou:2025tie,Wang:2025vfb, Cheng:2025lod}). Notably, a recent analysis by one of us~\cite{Wang:2025vfb} bypassed the linear equation of state parameterization entirely, instead reconstructing the DE density directly from DESI DR2 and CMB data. This approach yielded only a $1.2\sigma$ deviation from $\Lambda$CDM -- substantially weaker than the CPL-based claimed evidence. Such variations in significance between analysis procedures underscore the critical importance of understanding how prior assumptions shape cosmological inference, specifically in the context of the standard CPL parameterization which remains the focus of this work.

Among the possible explanations for the deviation from $\Lambda$CDM seen in the CPL analyses is DE from a \textit{quintessence}-like field~\cite{Peebles:1987ek,Ratra:1987rm}. However, constructing such a model with phantom behavior as preferred by DESI data presents theoretical challenges, requiring a violation of the null energy condition, which when sustained leads to unphysical scenarios, notably instabilities~\cite{Caldwell:1999ew,Hu:2004kh}. These difficulties can potentially be circumvented by coupling DE to other degrees of freedom, allowing the \textit{effective} equation of state to appear phantom while the underlying fields remain physically consistent~\cite{Das:2005yj,Carroll:2004hc,Khoury:2025txd,Huey:2004qv,Brax:2023qyp,Smith:2024ibv,Chen:2025ywv}. In such scenarios, the phantom behavior emerges only as an apparent effect from treating the coupled dark sector as a single component.

Given these theoretical difficulties in constructing physical models that reproduce the phantom-like behavior, combined with the marginal $3\sigma$ evidence for evolving DE and the parameterization sensitivity noted above, the choice of priors in these studies warrants careful scrutiny. Bayesian inference underpins cosmological data analysis, enabling constraints on physical models using observables such as the CMB~\cite{Planck:2018vyg} and large-scale structure (LSS) datasets~\cite{Troster:2019ean,Ivanov:2019hqk,DESI:2024mwx,Ivanov:2024jtl}. Central to this approach is the incorporation of a prior within Bayes' theorem, expressed as:
\begin{equation}
P(\theta | d, M) = \frac{\mathcal{L}(d | \theta, M) P(\theta|M)}{P(d | M)},
\end{equation}
where the posterior probability $P(\theta | d, M)$ represents the probability of model parameters $\theta$ given a model $M$ and observed data $d$. The likelihood $\mathcal{L}(d | \theta, M)$, which quantifies the probability of the data given specific parameter values, is combined with a carefully chosen prior distribution $P(\theta | M)$ to compute the posterior.\footnote{See~\cite{Trotta:2017wnx} for a full discussion of Bayesian inference in cosmology.} The evidence, $P(d | M)$, serves as a normalization constant, ensuring the posterior is a valid probability distribution, though it is less critical for parameter estimation alone.

\begin{figure}[!t]
    \centering
    \includegraphics[width=\linewidth]{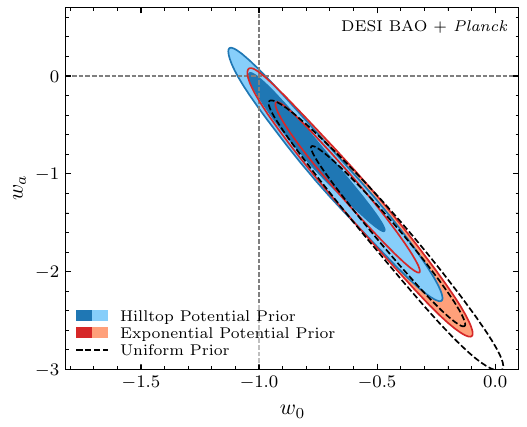}
    \caption{Two-dimensional posteriors in $w_0$ and $w_a$ at the 68\% and 95\% confidence levels from a joint fit to \textit{Planck} PR4 CMB anisotropies with lensing and DESI DR2 BAO under different priors for evolving dark energy. The gray dashed lines mark the $\Lambda$CDM limit of the $w_0w_a$CDM model, namely $(w_0,\,w_a)=(-1,0)$. For two degrees of freedom, the results are consistent with $\Lambda$CDM at the $\sim 1.3\sigma$ (hilltop, \textcolor{tabblue}{blue}) and $\sim 1.8\sigma$ (exponential, \textcolor{tabred}{red}) levels, in contrast with uniform priors (dashed black) that yield a $\sim 3.1 \sigma$ deviation. The figure highlights the impact of prior choices, with theory-informed priors that shift the inference toward $\Lambda$CDM, significantly reducing the nominal preference for evolving dark energy.}
    \label{fig:w0wa_cmb_bao}
\end{figure}

The influence of prior selection in Bayesian inference cannot be overstated when constraints are weak. While precise measurements with sharply peaked likelihoods are robust to changes in priors, marginal detections, such as the tentative evidence for evolving DE, are susceptible to prior-induced bias (see e.g.~\cite{Steinhardt:2025znn,Bayat:2025xfr}). The use of broad uniform priors on the CPL parameters $w_0$ and $w_a$, often mischaracterized as ``uninformative,'' represents a specific choice that influences the posterior distributions, especially in the context of DESI measurements. Although uniform priors are sometimes adopted in an attempt to minimize subjective influence, this choice is far from neutral in a Bayesian framework. By assigning equal probability to all parameter values, uniform priors implicitly assume that all regions of the $(w_0, w_a)$ space, including highly phantom-like regions, are as plausible as those resembling a cosmological constant. As we've discussed already, this point is problematic in the CPL model, as it disregards theoretical constraints and physical plausibility, potentially skewing the interpretation towards marginal detections.

To robustly address this consideration, one can incorporate \textit{theory-informed} priors when available. Since priors encode our degree of belief that a model will take on certain values, theoretical insights provide useful guidance. To address these concerns, we implement physically motivated priors derived from scalar field {\it quintessence} theories of DE via normalizing flows (NFs)~\cite{2015arXiv150505770J, 2017arXiv170507057P},  a class of machine learning algorithms which can learn complex probability distributions, 
focusing on hilltop and exponential potentials as two representative examples.
These priors incorporate theoretical expectations about viable DE models while remaining flexible enough to accommodate observational constraints. Our analysis, with results illustrated in Fig.~\ref{fig:w0wa_cmb_bao}, demonstrates that prior choices 
notably affect cosmological inference: the same DESI DR2 + \textit{Planck} data combination that yields a $\sim 3.1\sigma$ preference for dynamical DE under uniform $(w_0,w_a)$ priors falls to $\sim 1.3\sigma$  when physically motivated hilltop-potential priors are employed, with exponential-potential priors yielding an intermediate $\sim 1.8\sigma$ deviation. The two-dimensional posterior in the ($w_0$, $w_a$) plane reveals that the apparent preference for dynamical DE under uniform priors 
at least partially stems from regions of the parameter space that, while mathematically permissible, are physically disfavored or highly fine-tuned.

For completeness, we also analyze data combinations that include Type Ia supernovae, finding analogous shifts toward the $\Lambda$CDM limit, though with residual preferences that depend sensitively on the specific supernova compilation considered. Taken together, these findings highlight the critical role of prior assumptions in shaping cosmological inference and caution against over-interpreting marginal departures from $\Lambda$CDM that may arise when using uniform priors extending into regions of parameter space that are physically implausible.

This paper is organized as follows. In Sec.~\ref{ConThInfPrior} we discuss the construction of theory informed priors on the $(w_0,w_a)$ parameterization and our training of a NF to efficiently calculate posteriors. We next describe the Bayesian inference setup and datasets used in Sec.~\ref{AnalysisAndDatasets}. In Sec.~\ref{Results} we present the main results and we conclude in Sec.~\ref{DNC} with a discussion of their broader implications. 

\section{Constructing Theory-Informed Priors}\label{ConThInfPrior}

\noindent In this section, we describe our method for incorporating theory-informed priors from fundamental physics into observational analyses using the CPL parameterization. We focus on two representative classes of thawing quintessence: exponential and hilltop potentials. After introducing these models and their theory-informed priors, we present the mapping procedure to determine the corresponding distributions on $(w_0,w_a)$.  Finally, we describe the normalizing flow approach used to construct the theory-informed $(w_0,w_a)$ priors that can be efficiently applied to observational data without the need to directly resample the quintessence parameter space. 

\subsection{Quintessence Models and Theoretical Priors}

\noindent In the simplest realizations of quintessence, DE is driven by a single scalar field $\phi$  minimally
coupled to gravity. The evolution of this homogeneous field in a Friedmann-Lemaitre-Robertson-Walker (FLRW) universe is then governed by:
\begin{equation}
    \ddot{\phi} + 3H\dot{\phi} +
dV/d\phi
= 0, \label{eq:1}
\end{equation}
where dots represent derivatives with respect to cosmic time, $H$ is the Hubble expansion rate, and $V(\phi)$ is the potential associated with the scalar field, which at the background level behaves as a perfect fluid with density and equation of state parameter respectively equal to:
\begin{align}
    &\Omega_\phi=\frac{V(\phi)+\dot{\phi}^2/2}{3 H^2 M_{\rm pl}^2},
    \\
    &w_\phi=\frac{\dot{\phi}^2/2-V(\phi)}{\dot{\phi}^2/2+V(\phi)}, 
\end{align}
with $M_{\rm pl}=1/\sqrt{8\pi G}\approx 2.4\times 10^{18}\,{\rm GeV}$ being the reduced Planck mass. Thawing quintessence, which is the primary focus of this work, describes fields that remain frozen at early times due to Hubble friction dominating over the potential gradient in Eq.~\eqref{eq:1}. As the universe expands and $H$ decreases, the field {\it thaws} and begins rolling, acquiring kinetic energy that pushes the equation of state parameter $w$ above the initial value of $-1$. The present DE-dominated epoch corresponds then to this transitional phase between the frozen past and any eventual attractor.

As anticipated in the introduction, we examine here two representative potentials that capture different thawing behaviors. The first is the exponential potential:
\begin{equation}
    V_{\rm exp}(\phi)
=V_0e^{\lambda\,\phi/M_{\rm pl}}\label{eq:exp}
\end{equation}
where $V_0$ is the scale of the potential, which can be set to satisfy the Friedmann equations and reproduce the observed DE density today. Scalar fields with exponential potentials arise naturally in supergravity, modified gravity, and string theory constructions~\cite{Wetterich:1994bg,Binetruy:1998rz,Bedroya:2019snp}, and have been extensively studied as simple, well-motivated DE candidates. For $\lambda \sim \mathcal{O}(1)$ -- the parameter space of greatest theoretical interest, as we discuss below -- these potentials generically produce slowly thawing trajectories in which the equation of state $w$ increases gradually from $-1$ near the present epoch.

The second class of models we consider feature a hilltop potential:
\begin{equation}
    V_{\rm hill}(\phi)=V_0\left(1-\frac{k^2\phi^2}{2}\right)\label{eq:hilltop},
\end{equation}
where $k$ controls the potential steepness and $V_0$ is again fixed by the present-day DE density. Hilltops provide a good approximation to axion models with initial conditions near the top of a periodic potential, and more generally, to pseudo–Nambu–Goldstone boson scenarios in which the flatness near the maximum is protected by approximate shift symmetries~\cite{Dutta:2008qn,Frieman:1995pm}. The Hilltop model also emerges from a simple Higgs-like model $V(\Phi)= V_0 ( |\Phi|^2/v^2 -1)^2$ with $\phi \equiv |\Phi|$ and the hilltop regime realized in the limit $\phi \ll v$.

The steepness parameter $k$ controls how quickly $w$ departs from $-1$ once thawing begins: flatter hilltops lead to slower evolution, while steeper ones produce a more rapid rise in $w$ as the field accelerates.  The resulting variations in late-time evolution map to characteristic regions in the $(w_0,w_a)$ plane, making this class a useful complement to exponential potentials for building theory-informed priors.

For each potential, we identify the relevant model parameters and the considerations guiding our theoretical priors. For the exponential potential, the only free parameter is the slope $\lambda$. The apparent freedom in the initial field value $\phi_i$ deep in matter domination is in fact redundant due to the shift-rescaling symmetry of the model, namely for $\phi\rightarrow \phi+\phi_i$, $V_0\rightarrow V_0e^{\lambda \phi_i/M_{\rm pl}}$. Thus, we set $\phi_i=0$ without loss of generality, with $V_0$ that becomes fully fixed by requiring $\Omega_{\phi,0}=\Omega_{\rm DE,0}$. The parameter $\lambda$ is constrained from both theoretical and observational considerations. The de Sitter Swampland conjecture requires $|\nabla_\phi V|/V \gtrsim c \sim \mathcal{O}(1)$ in reduced Planck units for any consistent theory of quantum gravity with $V > 0$~\cite{Obied:2018sgi}, which for exponential potentials translates to $\lambda \gtrsim \mathcal{O}(1)$.\footnote{Note, the refined Swampland De Sitter conjecture~\cite{Agrawal:2018rcg} allows for the possibility that $V'/V=0$ if $V''<0$, however in the exponential model $V''>0$ at all times.} The requirement of a past radiation-dominated epoch followed by present-day acceleration imposes the bound $\lambda \lesssim \sqrt{3}$, which gets slightly looser, namely $\lambda \lesssim 2$~\cite{Andriot:2024jsh},  when allowing for observationally permitted spatial curvature. 

For the hilltop potential, we have two free parameters, the initial field value $\phi_i$ and the steepness parameter $k$, with $V_0$ again determined by matching the present-day DE density. As a low energy effective field theory (EFT), the model is directly constrained by the Swampland Distance Conjecture (SDC)~\cite{Ooguri:2006in}, namely that $|\Delta \phi|\equiv |\phi_0 - \phi_i| \ll M_{pl}$ in order for the EFT to be valid from the initial time to today. For $k \phi_i \ll 1$ the field is nearly stationary until the present time and we expect $\Delta \phi \ll 1/k$. The SDC then suggests that $k$ should not be much less than $M_{pl}^{-1}$.  
Indeed, analytically, the evolution of $\phi$ during dark-energy like (slow roll) evolution is given by $\Delta \phi \simeq \phi_i(e^{(k M_{pl})^2 \Delta N}-1)$ where $N$ is the number of e-folds of expansion. Since the field excursion is proportional to the initial value, $\phi_i$, and is exponentially suppressed by $k^2$, we infer that even for $\phi_i \ll M_{pl}$, and an ${\cal O}(1)$ number of e-folds of DE domination, the field excursion is subplanckian even for $k \gg M_{pl}^{-1}$.

The hilltop model can be further constrained by considering the UV completion, for example, either as an axion or an Higgs-like field. We first consider the case of an axion. Near their maximum, axion potentials of the form $V(\phi) \simeq V_0\cos(\phi/f)$ can be approximated as $V(\phi) \simeq V_0\left[1 - \phi^2/(2f^2)\right]$, matching our quadratic hilltop for $k \simeq 1/f$. This connection allows us to import theoretical priors from axion physics, where the string axiverse~\cite{Arvanitaki:2009fg} and explicit string compactifications~\cite{Broeckel:2021dpz,Halverson:2019cmy,Mehta:2020kwu,Mehta:2021pwf,Demirtas:2021gsq} predict log-uniform distributions for both axion masses and decay constants. General considerations of quantum gravity~\cite{Montero:2015ofa,Kallosh:1995hi,Banks:2003sx} and the Weak Gravity Conjecture~\cite{Arkani-Hamed:2006emk, Harlow:2022ich} all suggest axion decay constants are sub-Planckian, $f<M_{\rm pl}$ in any controlled effective field theory with a convergent instanton expansion. Additionally, the SDC, can be applied to axions~\cite{Baume:2016psm,Klaewer:2016kiy,Blumenhagen:2017cxt,Scalisi:2018eaz}, where it again predicts a breakdown of effective field theory for super-Planckian field excursions, and is again compatible with hilltop dark energy dynamics as described above.

The hilltop models can also be UV completed into a Higgs-like symmetry breaking potential with vacuum expectation value $v$. In this case we may identify $k\simeq 1/v$.  If the scalar is charged, the gauge field will eat the would-be axion and at low energies we can integrate it out, leaving only $\phi$ in the low energy EFT. While the model remains restricted by the SDC to $|\Delta \phi| < M_{\rm pl}$, this constraint is much weaker than requiring $v< M_{\rm pl}$. This is because for the field to still be evolving as DE today, i.e. slowly rolling down its potential,  we need $|\Delta \phi|\ll v$, making the SDC constraint easily satisfied for $k\lesssim 1/M_{\rm pl}$. This differs notably from the axion case where $k\simeq 1/f$ and the Weak Gravity Conjecture~\cite{Arkani-Hamed:2006emk, Harlow:2022ich} requires $f < M_{\rm pl}$, i.e. $k\gtrsim 1/M_{\rm pl}$, directly constraining the fundamental parameter.

Based on all of these considerations, we adopt broad flat uniform priors on $\lambda$, $\log_{10} k$ and $\log_{10}\phi_i$:
\begin{align}
    {\text{\it exponential potential}}:\quad &\lambda\sim\mathcal{U}(0,2); \label{eq:Prior_exp}\\
    {\text{\it hilltop potential}}: \quad &\log_{10}\left(k\cdot {M_{\rm pl}}\right)\sim \mathcal{U}(-4,\,1),\nonumber \\
    &\log_{10}\left(\phi_i/M_{\rm pl}\right)\sim \mathcal{U}(-7,\,-1). \label{eq:Prior_hilltop}
\end{align}
We deliberately choose priors that extend well beyond the theoretical limits discussed above, to ensure 
we capture the full range of phenomenologically viable parameter space. We expect that stricter theory priors would strengthen the results of this paper.\footnote{For the hilltop case, we also tested alternative prior choices, including uniform priors directly on $\phi_i/M_{\rm pl}$ up to $0.1$ as well as up to the Planck scale, and found close agreement with the results presented here, so we do not consider these alternative choices further.} 

\subsection{Mapping from Theory Parameters to $(w_0,\,w_a)$}

\noindent

It remains to translate the theory priors discussed above into corresponding distributions in the $(w_0,w_a)$ parameter space commonly used in observational analyses. This mapping is essential for understanding and identifying the
regions of the phenomenological parameter space that are naturally populated by theoretically well motivated models. 

The challenge in establishing this mapping stems from the fundamental difference between the physical evolution of $w_\phi(z)$ in quintessence models and the phenomenological CPL parameterization, Eq.~\eqref{eq:cpl}. While quintessence models produce smooth, physically motivated trajectories determined by field dynamics, Eq.~\eqref{eq:1}, the CPL form is merely a convenient fitting function, by construction linear in the scale factor. Following~\cite{Shlivko:2024llw,Shlivko:2025fgv}, we determine the correspondence by matching the expansion history rather than the equation of state directly, as the Hubble parameter $H(z)$ is what enters directly into cosmological observables.
Specifically, we search for the $(w_0,w_a)$ combination that best reproduces the Hubble evolution of a given quintessence model by minimizing the maximum relative error:\footnote{We note that the choice of minimizing $E_H$ is not unique. One could instead minimize other quantities, such as the comoving angular diameter distance 
$D_M$ directly, as well as adopt a different binning in redshift. These alternatives lead to only minor shifts in the induced $(w_0,w_a)$ distributions and do not affect the overall conclusions of our analysis.} 
\begin{equation}
    E_H=\underset{z<4}{\rm max}\left|\frac{H_{\rm CPL}(z) - H_\phi(z)}{H_\phi(z)}\right|,\label{eq:EH}
\end{equation}
where $H_\phi(z)$ is the Hubble evolution for the quintessence model under consideration, which we compute numerically using \texttt{class\_ede};\footnote{\url{https://github.com/mwt5345/class_ede}} $H_{\rm CPL}(z)$ corresponds to the analytical CPL parameterization~\cite{Chevallier:2000qy,Linder:2002et},
and $z<4$ roughly matches the redshift range probed by BAO and supernova observations.\footnote{Note, DESI BAO measurements span 7 redshift bins from $z=0.3$ to $z=2.33$, with the last bin extending out to $z\approx 3.5$~\cite{DESI:2024uvr,DESI:2024aqx}. Restricting our redshift range to more closely match these measurements (instead of taking $z$ all the way out to 4 as in Eq.~\eqref{eq:EH}) has effectively negligible impact on the results of our analysis.} As demonstrated in~\cite{Shlivko:2024llw}, this approach ensures accurate reproduction of key observables that directly enter BAO measurements, in particular the Hubble distance and angular diameter distance. However, for a given quintessence model there is no unique ``best-fit'' $(w_0,w_a)$ pair. Instead, there exists a region in the $(w_0,w_a)$ plane that reproduces the Hubble evolution with nearly equivalent accuracy, forming an approximate degeneracy where combinations satisfying $\Delta w_a/\Delta w_0 \approx -5$ produce comparably good matches to $H(z)$~\cite{Shlivko:2024llw}. This degeneracy reflects a fundamental limitation of the CPL parametrization, with its prescribed linear evolution in the scale factor, that cannot uniquely capture the distinct dynamics of the quintessence field evolution. 

To address this inherent ambiguity and construct meaningful prior distributions, we adopt a probabilistic approach. For each set of quintessence model parameters drawn from our theoretical priors, we generate a probability distribution over the $(w_0, w_a)$ plane weighted by the inverse square of the matching error, $P(w_0,w_a)\propto 1/E_H^{2}$.  This inverse-variance weighting naturally assigns higher probability to parameter combinations that better reproduce the quintessence expansion history. We evaluate this distribution over the range $w_0 \in [-3, 1]$ and $w_a \in [-3, 2]$, matching the bounds typically used in cosmological analyses.
\begin{figure}
    \centering
    \includegraphics[width=\linewidth]{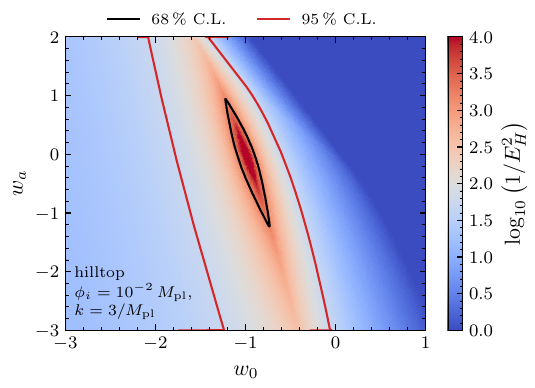}
    \caption{Inverse-variance weighted probability distribution $P(w_0,w_a)\propto 1/E_{H}^2$  showing the mapping from a hilltop quintessence model, Eq.~\eqref{eq:hilltop}, to the $(w_0, w_a)$ plane
    for a single model realization. Specifically, the example shown corresponds to $\phi_i=10^{-2}\,M_{\rm pl}$ and $k=3/M_{\rm pl}$,  with $H_0$ and $\Omega_m$ fixed to the {\it Planck} 2018 best-fit values~\cite{Planck:2018vyg} in order to compute $E_H$, the maximum relative error in matching the Hubble evolution defined in Eq.~\eqref{eq:EH}. The colorbar indicates the relative likelihood of each $(w_0,w_a)$ pair, shown as $\log_{10}(1/E_H^2)$, and the black and red contours enclose the 68\% and 95\% confidence levels (C.L.) of this distribution, respectively.  While the 68\% C.L. remains concentrated near $(w_0, w_a) \approx (-1, 0)$, the 95\% C.L. extends broadly across the parameter space, reflecting the degeneracy in matching the quintessence Hubble evolution with the CPL parameterization.}
    \label{fig:1}
\end{figure}

Fig.~\ref{fig:1} illustrates this probability distribution $P(w_0,w_a)$ for a representative set of hilltop model parameters, with an equivalent case for the exponential potential shown in Fig.~\ref{fig:A1} in the Appendix. The distribution exhibits several key features. The characteristic degeneracy direction is clearly visible, demonstrating that relatively good fits at the percent level can be achieved across a wide range of $w_a$ values. Despite this degeneracy, the 68\% confidence region remains concentrated in a relatively narrow area, with values close to $w_a \sim 0$ and $w_0\sim -1$ for this particular example. The 95\% confidence region, however, extends much more broadly, spanning nearly the entire range in $w_a$ while remaining mostly concentrated at $w_0 \gtrsim -1$. Notably, this broader region also includes $(w_0, w_a)$ pairs that reproduce the Hubble evolution with only 10-30\% accuracy.

To construct the overall theory-informed prior distribution on $(w_0, w_a)$, we must aggregate these probability distributions across all possible parameter values for a given quintessence model. We accomplish this by randomly sampling a single $(w_0, w_a)$ pair from within the 95\% confidence region of this distribution. This choice is deliberately conservative -- given that current BAO measurements achieve percent-level precision, parameter combinations that match $H(z)$ with only 30\% accuracy are poor representatives of the underlying field dynamics, yet we include them to avoid artificially narrowing the induced $(w_0, w_a)$ distribution. This approach ensures our theory-informed priors remain broad and do not exclude regions that, while imperfect matches to the quintessence dynamics, still fall within the conventional parameter space explored in observational analyses. By repeating this procedure across the full range of model parameters sampled from our theoretical priors, we construct the induced distributions in the $(w_0, w_a)$ space that reflect the collective predictions of each class of quintessence models. The implementation of the details of this mapping, including the treatment of $\Lambda
$CDM parameters necessary for constructing reliable conditional distributions, are discussed in the following section.

\subsection{Obtaining $(w_0, w_a)$ Priors with Normalizing Flows}

\begin{figure*}
    \centering
    \includegraphics[width=.95\linewidth]{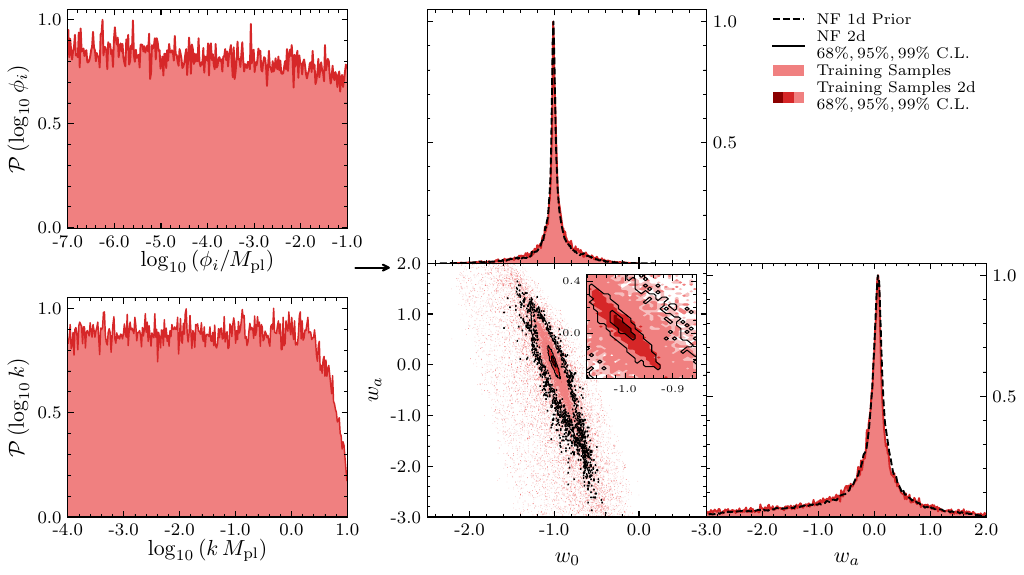}
    \caption{Theory-informed prior distributions for the hilltop quintessence model, Eq.~\eqref{eq:hilltop}, obtained through normalizing flow (NF) training. {\it Left panel:} Model parameter samples drawn from broad flat priors, showing $\phi_i/M_{\rm pl}$ ({\it top}) and $k$ ({\it bottom}) in light red histograms, confirming full coverage of the theoretical prior ranges. {\it Right panel: } Triangle plot displaying the mapped $(w_0, w_a)$ distributions. Light red shows the training samples obtained through the matching procedure, with increasingly darker shades indicating the 99\%, 95\%, and 68\% confidence levels for the 2D distribution. Black lines represent the learned NF priors: dashed lines for the 1D marginalized distributions and solid contours for the 2D joint distribution (99\%, 95\%, and 68\% C.L.). The inset within the 2D joint distribution panel provides a zoomed view of the 68\% and 95\% contours, demonstrating the excellent agreement between the NF output and training data. The strongly peaked structure near $(w_0, w_a) \approx (-1, 0)$ with extended tails reveals how, minimally-biased, flat priors on fundamental model parameters naturally induce non-uniform priors on the phenomenological $(w_0, w_a)$ space,  potentially altering cosmological inferences significantly.}
    \label{fig:NF_priors}
\end{figure*}

\noindent 

To efficiently generate theory-informed priors on $(w_0, w_a)$ that can be applied to existing observational analyses, we follow~\cite{Toomey:2024ita} and employ normalizing flows (NFs), a class of generative machine learning algorithms that learn bijective mappings between complex target distributions and simple base distributions~\cite{nfs}. In our context, we use a conditional NF~\cite{2015arXiv150505770J, 2017arXiv170507057P}
 to learn theory-informed priors on the DE equation-of-state parameters $(w_0, w_a)$   trained on outputs from flat priors over the fundamental quintessence parameters (for either the hilltop or exponential potential) together with variations in $(\omega_c,\,\omega_b, \,H_0)$. The NF models the conditional density $P_{\mathrm{th}}(w_0, w_a \mid \omega_c, \omega_b, H_0)$, providing a smooth, non-Gaussian, and highly flexible representation of the theory predictions. 
 
 In what follows, we outline how the NF is trained on simulated quintessence realizations, how it is used to construct smooth distributions in $(w_0, w_a)$, and finally how these learned priors are incorporated into MCMC analyses of real data. For a given set (choice) of theory parameters, drawn from our
theoretical model priors, Eq.~\eqref{eq:Prior_exp} and~\eqref{eq:Prior_hilltop}, as well as cosmological parameters, the approach of the previous section produces a five dimensional data set ($w_0$, $w_a$, $\omega_b$, $\omega_c$, $H_0$) for that particular choice of parameters. We repeat this procedure for a large number (50,000) of sets of parameters. However, these discrete samples are impractical to use directly, and re-performing this procedure for every MCMC sample is computationally inefficient. We therefore train an NF on these 50,000 realizations to learn the smooth conditional probability distribution $P_{\mathrm{th}}(w_0, w_a \mid \omega_c, \omega_b, H_0)$ by minimizing the loss function (see Appendix \ref{sec:A2} for more details). The learned  probability distribution can subsequently be used as a prior when analyzing real data (in the following section). In practice, we generate MCMC chains using the uniform priors on $(w_0,\, w_a)$ from the DESI analyses and post-process them by drawing from the prior probability distribution for $(w_0,\,w_a)$ obtained with the NF. This approach yields two major computational advantages: (i) it avoids the need to sample directly in the theory parameters and perform the matching procedure described in the previous section at every MCMC step, and (ii) it allows existing analyses to be updated efficiently by reweighting the chains with the learned theory-informed priors rather than re-running the full parameter inference.

Before presenting the results, we note that the $\Lambda$CDM parameters are drawn from broad uniform priors matching those used in the DESI analyses~\cite{DESI:2025zgx,DESI:2025fii}, ensuring consistency. The complete training dataset comprises 5 parameters (2 CPL + 3 $\Lambda$CDM) plus quintessence-specific parameters -- two additional parameters ($k$ and $\phi_i$) for the hilltop model and one ($\lambda$) for the exponential model -- that we marginalize over.

Fig.~\ref{fig:NF_priors} displays the learned marginalized distributions for the  $(w_0, w_a)$ parameters in the hilltop case, with the corresponding exponential case shown in Fig.~\ref{fig:A2} in the Appendix.  The figure illustrates a 2D slice in $w_0, w_a$ of the sampled five dimensional data set ($w_0$, $w_a$, $\omega_b$, $\omega_c$, $H_0$) for hilltop quintessence (and Fig.~\ref{fig:A2} the same for exponential quintessence).   
The figure also shows our training data, both the original samples in quintessence parameter space, confirming full coverage of the prior ranges, and the mapped $(w_0, w_a)$ pairs obtained through the probabilistic matching procedure described in the previous section. Two key features emerge from this analysis. First, the normalizing flow output (black solid lines) accurately reproduces the distribution of mapped $(w_0, w_a)$ pairs from our training set, validating the NF's ability to learn this complex mapping. Second, and most significantly, the induced distributions on $(w_0, w_a)$ are decidedly non-uniform. They exhibit strong peaks near the $\Lambda$CDM values of $w_0= -1$ and $w_a = 0$, with extended tails particularly prominent in the negative $w_a$ direction. These qualitative features persist for the exponential potential, Fig.~\ref{fig:A2}, though with a somewhat broader peak reflecting its slower thawing dynamics. 

Thus we see that hilltop and exponential potentials (in the bounded regimes of parameter space as discussed above, in Eqs.~\eqref{eq:Prior_exp} and~\eqref{eq:Prior_hilltop}) produce a probability distribution in $(w_0,\, w_a)$ peaked at the $\Lambda$CDM limit in the CPL parametrization, although individual parameter choices within these models can yield cosmologies far from $\Lambda$CDM.

These result carry important implications for cosmological analyses. The peaked structure of the theory-informed priors emerged naturally from broad, minimally-biased priors on the fundamental model parameters, where we only imposed theoretically motivated boundaries and chose between linear or logarithmic sampling based on physical arguments. The transformation from these flat priors in model space to peaked priors in $(w_0, w_a)$ space demonstrates that uniform priors on $(w_0, w_a)$ -- often treated as ``uninformative'' -- are in fact highly informative: they allocate substantial prior volume to regions of the $(w_0,w_a)$ plane that are not representative of realistic quintessence dynamics, thereby skewing parameter inferences and potentially overstating evidence for evolving DE. For example, our theory-informed priors disfavor large negative for $w_0$ and $w_a$, with zero support for $(w_0,w_a)\lesssim (-2,-2)$ as seen in Figs.~\ref{fig:NF_priors} and~\ref{fig:A2}. This is in contrast to the uniform priors used in the DESI analyses which assign equal probability to this unphysical region, down to $(-3,-3)$, and the $\Lambda$CDM point at $(-1, 0)$. To quantify how these prior biases affect cosmological inferences, we reanalyze current datasets in a later section with the learned NF priors on $(w_0,w_a)$ developed here.

For clarity, Fig.~\ref{fig:NF_priors} displays only the marginalized 
 priors;  in practice, when post-processing the chains from current datasets, we employ the full NF priors conditioned on the relevant $\Lambda$CDM parameters, 
 at each sample point. This conditioning enables fast evaluation and sampling while preserving physical correlations and, crucially, eliminates the need for a full, time-intensive reanalysis with updated priors.
We provide further details on the NFs implementation and training in Appendix~\ref{sec:A2}.

\section{Analysis \& Data sets} \label{AnalysisAndDatasets}

 \begin{figure*}[!t]
  \centering
    \includegraphics[width=\linewidth]{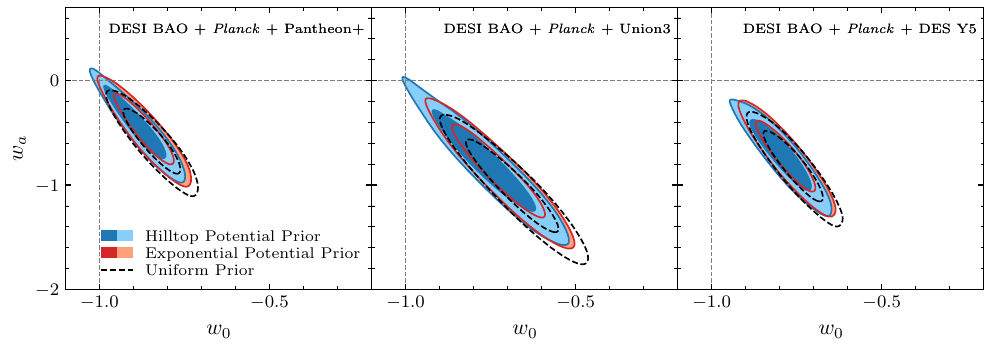}

  \caption{Two-dimensional posteriors in $w_0$ and $w_a$ at the 68\% and 95\% confidence levels from a joint fit to \textit{Planck} PR4 CMB anisotropies with lensing, DESI DR2 BAO, and Type Ia supernovae samples, under different priors for evolving dark energy, in their corresponding color and line-style. Specifically, the panels show the Pantheon+ ({\it left}), Union3 ({\it middle}), and DES Y5 ({\it right}) SNe Ia samples. The gray dashed lines mark the $\Lambda$CDM limit at $(w_0,\,w_a)=(-1,\,0)$. Across all three data combinations, theory-informed priors shift the inference toward $\Lambda$CDM and reduce the marginal preference for evolving dark energy.}
  \label{fig:three_pdfs}
\end{figure*}
Before examining our observational constraints in detail, it is instructive to consider a fundamental statistical challenge that affects both Bayesian and frequentist analyses of DE models within the CPL parametrization. As demonstrated in the previous section, many ($w_0,\,w_a$) combinations yield essentially the same expansion history (see e.g. Fig.~\ref{fig:1}), resulting in a likelihood surface that is nearly flat along this degeneracy direction when fitted to observational data. Therefore, extending the $\Lambda$CDM model with two additional CPL parameters increases the dimensionality of the parameter space  without substantially improving the model's ability to fit the data, since movements along the degeneracy direction leave the likelihood nearly unchanged. 

As a result, even modest reductions in $\chi^2$ can arise purely from statistical fluctuations in the likelihood as the model explores this expanded parameter volume, rather than from capturing genuine physical effects. In this context, profile likelihood analyses, often viewed as prior-independent, are also subject to the same pitfalls as uniform–prior Bayesian analyses as both approaches can inflate the apparent preference for evolving DE because of parameter–volume effects and likelihood fluctuations. This motivates the need for physically motivated priors that suppress implausible regions of parameter space and yield more robust inferences about the nature of DE.

With this statistical caveat in mind, we now turn to our analysis. We combine two dataset configurations:  (1) \textit{Planck} CMB data with DESI DR2 BAO measurements; (2) this baseline supplemented with Type Ia supernovae likelihoods.
For CMB constraints, we use the \textit{Planck} PR4 data release, incorporating NPIPE-based high- and low-$\ell$ temperature and polarization likelihoods, along with the PR4 lensing likelihood~\cite{Efstathiou:2019mdh,Rosenberg:2022sdy,Carron:2022eyg}. These datasets provide the most consistent and robust \textit{Planck} measurements, with improved foreground mitigation and noise modeling compared to the legacy release.
For BAO constraints, we employ the latest DESI DR2, which provides the most precise  measurements across a wide redshift range. For Type Ia supernovae constraints, we utilize three independent compilations: Pantheon+~\cite{Scolnic:2021amr,Brout:2022vxf}, Union3~\cite{Rubin:2023jdq}, and DES Year 5 (hereafter DES Y5)~\cite{DES:2024jxu}, enabling us to assess the robustness of our results across different samples. All likelihoods are implemented using the \texttt{Cobaya} framework~\citep{Torrado:2020dgo} with default settings for the supernova analyses.

For modeling cosmological observables, we employ the default CPL implementation in the Einstein-Boltzmann code \texttt{CLASS}~\cite{Blas:2011rf} with perturbations for likelihood evaluation. Following~\cite{DESI:2025zgx}, we adopt broad uniform priors for our base analysis over the parameter set $(\omega_{\rm c}, \,\omega_{\rm b},\, 100\theta_{\rm MC},\, \log(10^{10}A_{\rm s}),\, n_{\rm s},\, \tau,\, w_0,\, w_a)$, comprising the six standard $\Lambda$CDM parameters plus the two CPL parameters. Specifically, we impose $w_0 \in \mathcal{U}[-3,1]$ and $w_a \in \mathcal{U}[-3,2]$. We sample posterior distributions with Markov chain Monte Carlo (MCMC) using the Metropolis-Hastings algorithm~\cite{2005math......2099N}, running chains until achieving Gelman-Rubin convergence~\cite{10.1214/ss/1177011136} of $R - 1 < 0.01$. We report one-dimensional posteriors as the mean with 68\% minimum credible intervals, analyzing all MCMC chains with \texttt{GetDist}~\cite{Lewis:2019xzd}. Using our NF, implemented in \texttt{pytorch}~\cite{2019arXiv191201703P} and utilizing the \texttt{nflows}~\cite{nflows} library, we construct the constraints under the theory priors by post-processing our uniform prior samples with \texttt{GetDist}.

\section{Updated CMB, BAO constraints with physical prior}\label{Results}

\begin{table*}[!t]
	\centering
	\subfloat[Uniform Prior on $(w_0,w_a)$]{%
		\begingroup
		\renewcommand{\arraystretch}{1.2}
		\begin{tabular}{@{}l@{\hspace{1em}}c@{\hspace{1em}}c@{\hspace{1em}}c@{\hspace{1em}}c@{}}
			\toprule
			Dataset & $\Omega_m$ & $H_0$ [km s$^{-1}$ Mpc$^{-1}$] & $w_0$ & $w_a$ \\
			\midrule[0.065em]
			DESI BAO + {\it Planck}  & $0.350\pm 0.022$ & $63.8^{+1.8}_{-2.1}$ & $-0.45\pm 0.21$ & $-1.64\pm 0.59$ \\
			DESI BAO + {\it Planck}  + Pantheon+ & $0.311\pm 0.006$ & $67.5\pm 0.6$ & $-0.84\pm 0.06$ & $-0.59^{+0.21}_{-0.19}$\\
			DESI BAO + {\it Planck}  + Union3 & $0.327\pm 0.009$ & $66.0\pm 0.9$ & $-0.68\pm 0.09$ & $-1.03^{+0.30}_{-0.27}$\\
			DESI BAO + {\it Planck}  + DES Y5 & $0.319\pm 0.006$ & $66.7\pm 0.6$ & $-0.76\pm 0.06$ & $-0.83^{+0.23}_{-0.21}$\\
			\bottomrule
		\end{tabular}
		\endgroup
	}
	
	\vspace{1pt}
	
	\subfloat[Exponential Potential Prior on $(w_0,w_a)$]{%
		\begingroup
		\renewcommand{\arraystretch}{1.2}
		\begin{tabular}{@{}l@{\hspace{1em}}c@{\hspace{1em}}c@{\hspace{1em}}c@{\hspace{1em}}c@{}}
			\toprule
			Dataset & $\Omega_m$ & $H_0$ [km s$^{-1}$ Mpc$^{-1}$] & $w_0$ & $w_a$ \\
			\midrule[0.065em]
			DESI BAO + {\it Planck}  & $0.333^{+0.016}_{-0.024}$ & $65.4^{+2.2}_{-1.6}$ & $-0.63^{+0.15}_{-0.25}$ & $-1.12^{+0.68}_{-0.43}$\\
			DESI BAO + {\it Planck}  + Pantheon+ & $0.311\pm 0.006$ & $67.5\pm 0.6$ & $-0.87\pm 0.06$ & $-0.46\pm 0.22$ \\
			DESI BAO + {\it Planck}  + Union3 & $0.323\pm 0.009$ & $66.2\pm 0.8$ & $-0.73\pm 0.09$ & $-0.87\pm 0.29$\\
			DESI BAO + {\it Planck}  + DES Y5 & $0.318\pm 0.006$ & $66.8\pm 0.6$ & $-0.78\pm 0.06$ & $-0.73^{+0.25}_{-0.22}$\\
			\bottomrule
		\end{tabular}
		\endgroup
	}
	
	\vspace{1pt}
	
	\subfloat[Hilltop Potential Prior on $(w_0,w_a)$]{%
		\begingroup
		\renewcommand{\arraystretch}{1.2}
		\begin{tabular}{@{}l@{\hspace{1em}}c@{\hspace{1em}}c@{\hspace{1em}}c@{\hspace{1em}}c@{}}
			\toprule
			Dataset & $\Omega_m$ & $H_0$ [km s$^{-1}$ Mpc$^{-1}$] & $w_0$ & $w_a$ \\
			\midrule[0.065em]
			DESI BAO + {\it Planck}  & $0.318^{+0.009}_{-0.021}$ & $66.9^{+2.1}_{-1.0}$ & $-0.78^{+0.09}_{-0.22}$ & $-0.74^{+0.61}_{-0.30}$ \\
			DESI BAO + {\it Planck}  + Pantheon+ & $0.308\pm 0.005$ & $67.8^{+0.6}_{-0.5}$ & $-0.89\pm 0.06$ & $-0.43\pm 0.22$\\
			DESI BAO + {\it Planck}  + Union3 & $0.320\pm 0.009$ & $66.7\pm 0.9$ & $-0.76^{+0.09}_{-0.10}$ & $-0.80\pm 0.32$ \\
			DESI BAO + {\it Planck}  + DES Y5 & $0.316\pm 0.006$ & $67.0\pm 0.6$ & $-0.79\pm 0.06$ & $-0.72^{+0.25}_{-0.21}$\\
			\bottomrule
		\end{tabular}
		\endgroup
	}
	
	\vspace{-5pt}
	
	\caption{Summary table of key cosmological parameter constraints from DESI DR2 BAO in combination with {\it Planck} and additional supernova datasets, under different priors, in the $w_0 w_a$CDM model. Results quoted are the marginalized posterior means and their corresponding 68\% credible intervals, specifically for $w_0$, $w_a$, $H_0$ and $\Omega_m$. Together, these results show that theory-informed priors shift $(w_0,w_a)$ toward $(-1,0)$, reducing the apparent preference for evolving dark energy. See Tab.~\ref{tab:analysis_all} in the appendix for the complete set of parameter constraints.}
	\label{table_post}
\end{table*}

\noindent In this section, we investigate how theory-informed priors affect the constraints on evolving DE within the CPL parameterization. We employ our trained normalizing flow (NF) to reweight the MCMC chains originally generated with uniform priors on $w_0$ and $w_a$. Our analysis encompasses the primary results from the DESI DR2 analysis~\cite{DESI:2025zgx, DESI:2025fii}, combining \textit{Planck} CMB and DESI DR2 BAO measurements both independently and in conjunction with Type Ia supernovae data from the three different samples: Pantheon+, Union3, and DES Y5.

Our main result is shown in Fig.~\ref{fig:w0wa_cmb_bao}, which presents the posterior distributions in the ($w_0,\,w_a$) plane from a joint analysis of \textit{Planck} and DESI DR2 data under different prior choices, and in Tab.~\ref{table_post}, which reports the corresponding numerical constraints. Specifically, this figure and table illustrate our results for three different cases: the uniform prior used by DESI, and our two theory-motivated priors. If we use the same uniform priors on $w_0$ and $w_a$ as the DESI collaboration, our analysis reiterates their $\sim 3.1\sigma$ preference for evolving DE over $\Lambda$CDM and the following CPL parameter constraints:
\begin{equation}
\left.
\begin{aligned}
w_0 &= -0.45\pm 0.21   \\
w_a &= -1.64\pm 0.59 
\end{aligned}
\ \right\} \quad \substack{\text{DESI BAO + {\it Planck} } \\ \text{uniform prior}}~,
\end{equation}
and further demonstrates their evidence for a phantom crossing. However, the point of this paper is to argue that the choice of uniform priors is ill-suited for this parametrization.

To assess the sensitivity of this result to prior assumptions, we employ our trained NF to implement theory-informed priors for $w_0$ and $w_a$ through post-processing of the uniform-prior MCMC samples. We first examine the impact of the less restrictive physical prior derived from the exponential quintessence model -- as evidenced by its larger prior volume. Under this prior, the preference for CPL over $\Lambda$CDM diminishes to $\sim 1.8\sigma$, i.e. there is consistency with $\Lambda$CDM at the $\sim 1.8\sigma$ level, with the corresponding constraints shifting to:
\begin{equation}
\left.
\begin{aligned}
w_0 &= -0.63^{+0.15}_{-0.25}    \\
w_a &= -1.12^{+0.68}_{-0.43} 
\end{aligned}
\ \right\} \quad \substack{\text{ DESI BAO + {\it Planck}}  \\ \text{exponential potential prior}}~.
\end{equation}
This represents a substantial modification of the posterior distributions solely due to the change in prior assumptions.

The effect becomes even more pronounced when applying priors from the hilltop model. In this case, the result is consistent with $\Lambda$CDM at $\sim 1.3\sigma$, with the marginalized CPL posteriors given by:
\begin{equation}
\left.
\begin{aligned}
w_0 &=-0.78^{+0.09}_{-0.22}  \\
w_a &= -0.74^{+0.61}_{-0.30}
\end{aligned}
\ \right\}  \quad \substack{\text{ DESI BAO + {\it Planck} } \\ \text{hilltop potential prior}}~,
\end{equation}
which have in fact significantly shifted toward the $\Lambda$CDM limit of $(w_0,\,w_a)=(-1,\,0)$. Collectively, our results demonstrate that when theory-informed priors on $w_0$ and $w_a$ are employed, the evidence for evolving DE over $\Lambda$CDM is substantially weakened, suggesting that the original detection 
is at least in part driven by prior assumptions rather than the data itself. In addition, these findings are broadly consistent with dedicated analyses sampling directly the quintessence models' parameters~\cite{Payeur:2024dnq, Wolf:2024eph, Ramadan:2024kmn, Akrami:2025zlb}, further demonstrating the power of the NF approach to efficiently re-weight existing MCMC chains without the need to rerun full parameter inference.

For completeness, we also perform a joint analysis of CMB, BAO, and Type Ia supernovae (SNe Ia) data, where the inclusion of SNe Ia constraints complements the CMB and BAO combination, effectively reducing the $w_0$--$w_a$ degeneracy shown in Fig.~\ref{fig:w0wa_cmb_bao}. Cosmological parameter constraints are summarized in Tab.~\ref{table_post}, with posterior distributions presented in Fig.~\ref{fig:three_pdfs} (see also Tab.~\ref{tab:analysis_all} in the appendix for the complete set of constraints). In line with previous trends, physical priors shift the posteriors toward the $\Lambda$CDM model, leaving the data combinations including Pantheon+ and Union3 SNe Ia consistent with $\Lambda$CDM at roughly the $\sim 1\sigma$ and $\sim 2\sigma$ level, respectively. The shift for the DES Y5 sample is less dramatic, but still results in an appreciable movement of the posterior towards the $\Lambda$CDM limit. We note that these shifts are reflected in the constrained parameter values reported in Tab.~\ref{table_post}, with 
($w_0,\,w_a$) moving toward ($-1,\,0$) for all dataset combinations. In particular, for the DESI BAO + {\it Planck}  combination we also see $H_0$
and $\Omega_m$ shifting upward and downward, respectively, toward their baseline $\Lambda
$CDM values, namely $\Omega_m=0.303 \pm 0.004$ and $H_0=68.2 \pm 0.3$ from~\cite{DESI:2025zgx}.

While the inclusion of SNe Ia data introduces a residual preference for evolving DE, in particular when the DES Y5 sample is considered, readers should exercise caution in interpreting these combined results. To start, as further highlighted in our analysis, different SNe Ia compilations yield significantly different cosmological constraints despite the substantial overlap in their samples, raising concerns that systematic effects rather than genuine cosmological signals may be at play.  The DES Y5 sample, which produces the strongest hint for DE evolution, has been particularly scrutinized in this context. For instance, Ref.~\cite{Efstathiou:2024xcq} identified a possible $\sim 0.04$ magnitude calibration offset in the low-redshift supernovae common to DES~Y5 and Pantheon+ that, once corrected, brings DES Y5 into good agreement with $\Lambda$CDM (see also~\cite{Huang:2025som}). In response, the DES team carried out a simulation-based re-analysis and argued that these offsets are understood and properly accounted for within their systematic error budget~\cite{DES:2025tir}. Nonetheless, several independent studies~\cite{Gialamas:2024lyw, Notari:2024zmi,DESI:2025zgx,Wang:2025bkk,Huang:2025som, Afroz:2025iwo,  Teixeira:2025czm,Chaudhary:2025vzy} have highlighted that the apparent preference for evolving DE originates almost entirely from the lowest-redshift ($z<0.1$) supernovae, which are largely external to the DES survey itself. When these data are excluded, the DES Y5 preference for DE evolution effectively disappears, with the actual DES supernovae at $z>0.1$ tending to pull the combined constraints toward rather than away from $\Lambda$CDM~\cite{Gialamas:2024lyw, Notari:2024zmi, Wang:2025bkk,Huang:2025som, Chaudhary:2025vzy, Guedezounme:2025wav}.

Taking all of these considerations into account, the most robust cosmological inference to date remains  that drawn from the DESI BAO and CMB combination alone (see also Ref.~\cite{Cortes:2025joz} for a related statistical perspective) which, as we have shown here, provides essentially no evidence for evolving DE once theory-informed priors are imposed.

Lastly, even if one were to take the combined constraints including SNe Ia samples at face value, it is worth noting that the marginalized posteriors for $(w_0,\,w_a)$ remain noticeably displaced from the regions favored by our theory-informed priors, which are representative of minimal quintessence models. This suggests that, if the deviations are not simply the result of unaccounted systematics, they may instead be pointing toward a more complex dark sector, in which the apparent phantom-like behavior of the effective equation of state arises from couplings between DE and additional degrees of freedom, as has been explored in a variety of scenarios in the literature~\cite{Ye:2024ywg,Chakraborty:2024xas,Wang:2024hks,Giare:2024smz,Li:2024qso,Aboubrahim:2024cyk,Pan:2025psn,Cai:2025mas, Wolf:2024stt,Ye:2024zpk,Tiwari:2024gzo,Chakraborty:2025syu,Khoury:2025txd,Wolf:2025jed,Bedroya:2025fwh,Brax:2025ahm, Li:2025owk,Sabogal:2025mkp,Tsedrik:2025cwc,Zhai:2025hfi,Shah:2025ayl,Silva:2025hxw,Pan:2025qwy,Yashiki:2025loj,Barman:2025ryg,Li:2025ula, Wang:2025znm,Petri:2025swg, vanderWesthuizen:2025rip}. Even in such cases, however, it would remain essential to carry out theory-informed analyses to assess the robustness of the inferred departures from $\Lambda$CDM.

\section{Discussion \& Conclusion}\label{DNC}

\noindent In this work we have studied the impact of prior choices on the evidence for evolving DE, as suggested by recent DESI BAO results~\cite{DESI:2024mwx,DESI:2025zgx}. Specifically, we map two representative quintessence models -- hilltop and exponential -- into the effective CPL parameterization. Using a normalizing flow  to construct the corresponding theory-informed priors on $(w_0,w_a)$, we find that the apparent $\sim 3.1\sigma$ evidence for evolving DE in DESI DR2 BAO + \textit{Planck} data is reduced to $1.8\sigma$ and $1.3\sigma$ under exponential and hilltop potential priors, respectively, rendering the results statistically consistent with $\Lambda$CDM. When supernova data are included (with the caveat that systematics may be significant), the preference for evolving DE is also weakened compared to the DESI DR2 claims, with the $(w_0,\,w_a)$ posteriors shifting closer to the $\Lambda$CDM limit of $(-1,0)$. The only case in which  a residual preference for evolving DE at the $\sim 3\sigma$ level remains is the DES Y5 sample, reduced from claims of 4$\sigma$ significance, though it is not yet clear whether this signal reflects new physics  or simply residual systematics in the low-redshift supernova sample (see e.g.~\cite{Efstathiou:2024xcq,Notari:2024zmi,DES:2025tir, Cortes:2025joz}). 

As a whole, our analysis underscores the profound influence of prior choice on cosmological parameter inference when \textit{observational evidence for deviations from $\Lambda$CDM is marginal}. As demonstrated in Sec.~\ref{Results}, the choice of prior on the CPL parameters $w_0$ and $w_a$ can substantially  alter the perceived significance of evolving DE, with uniform priors -- often mischaracterized as ``uninformative'' -- potentially introducing bias by assigning equal probability to regions of the parameter space that are physically disfavored. This effect is especially pronounced in cases of weak detections, where the likelihood is not sharply peaked, as seen in the DESI DR2 analysis combined with \textit{Planck} CMB and Type Ia SNe data.

The source of this issue lies in the phenomenological nature of the CPL ansatz 
which in turn lacks intrinsic physical scales that dictate a natural scale or dynamic range  for $(w_0,w_a)$. Assigning uniform probability across all values  
therefore embeds implicit assumptions 
about the plausibility of phantom-like dynamics. Similar issues arise in other phenomenological models, such as early dark energy parameterizations, where unphysical priors can also bias the derived costraints~\cite{Toomey:2024ita}. Our results confirm that even flat priors 
can shape cosmological inference when likelihoods are broad, unlike in the case of sharply peaked, high-significance detections where prior dependence is reduced.

Within this context, frequentist approaches do not automatically cure these issues.
Profile-likelihood analyses have been shown to mirror the DESI preference for evolving DE~\cite{Herold:2025hkb}, however this should not be taken to mean that the DESI results are robust to prior-volume effects, as further evidenced by our analysis. In fact, when the likelihood has flat directions bounded by parameter limits, profiling can ``chase'' the best-fitting corners along the long degeneracy directions (e.g. the strongly phantom region in the scenario considered here), yielding modest improvements in $\Delta\chi^2$ that reflect the model flexibility rather than a data-driven signal. In such cases, the usual conversion of 
$\Delta\chi^2$ to a nominal $\sigma$ significance can overstate the evidence, and agreement between uniform-prior posteriors and profile likelihoods does not by itself establish robustness. Using a Bayesian framework with theory motivated priors is therefore essential to distinguish true signals from artifacts of parameterization and prior volume, providing a transparent way to restrict the parameter space to physically plausible regions and test whether the preference persists.

As a practical point, we emphasize that our NF-based framework for post-processing MCMC chains represents not only a highly accurate method for exploring a range of models (see Figs.~\ref{fig:NF_priors} and \ref{fig:A1}),  but also a highly efficient one, requiring minimal extra compute time on top of our base chains -- in the models explored here the total extra compute time was $\mathcal{O}(1)$~hour total CPU time. This enhanced efficiency opens the possibility for scalable and robust cosmological inference across a range on theoretical models.

In conclusion, our reanalysis of DESI DR2 data with theory-informed priors demonstrates that the evidence for evolving DE is weaker than suggested by uniform-prior analyses. What appears to be a $>3\sigma$ deviation from $\Lambda$CDM is, under physically motivated priors, statistically consistent with the concordance model. This underscores the importance of carefully assessing prior dependence in cosmological inference, particularly when interpreting results that could otherwise be taken as indications of physics beyond $\Lambda$CDM.

\acknowledgments
The authors thank Ellie Hughes and Mikhail Ivanov for useful conversations.
This material is based upon work supported by the U.S. Department of Energy, Office of Science, Office of High Energy Physics of U.S. Department of Energy under grant Contract Number  DE-SC0012567. MWT  acknowledges financial support from the Simons Foundation (Grant Number 929255).
K.F. and G.M. are grateful for support from
the Jeff \& Gail Kodosky Endowed Chair in Physics at the University of Texas. 
K.F. and G.M. also acknowledge
support from the Swedish Research Council (Contract
No. 638-2013-8993).   E.M. is supported in part by a Discovery Grant from the Natural Sciences and Engineering Research Council of Canada, and by a New Investigator Operating Grant from Research Manitoba.

\bibliographystyle{apsrev4-1}
\bibliography{bibl}

\pagebreak
\clearpage
\widetext
\appendix

\section{Normalizing Flow Architecture \& Training}\label{sec:A2}

\begin{figure}[!h]
    \centering
    \includegraphics[width=0.5\linewidth]{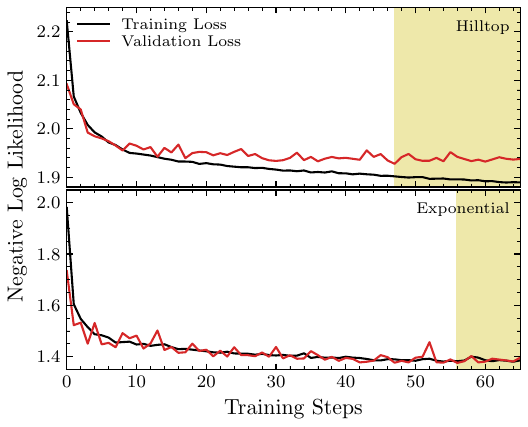}
    \caption{The training (black) and validation (\textcolor{tabred}{red}) loss for our $\Lambda$CDM parameter conditioned normalizing flow for hilltop ({\it top panel}) and exponential ({\it bottom panel}) priors. The \textcolor{goldenrod}{yellow} shaded region marks the interval after the minimum validation loss was reached before early stopping was triggered.}
    \label{fig:nf_loss}
\end{figure}

We construct a theory-informed prior on the dark energy equation-of-state parameters $(w_0, w_a)$ using a conditional normalizing flow (NF)~\cite{2015arXiv150505770J, 2017arXiv170507057P} trained on outputs from a theoretical model in which $(\omega_c, \omega_b, H_0)$ are also varied. The NF models the conditional density $P_{\mathrm{th}}(w_0, w_a \mid \omega_c, \omega_b, H_0)$, providing a smooth, non-Gaussian, and highly flexible representation of the theory predictions. This learned prior is subsequently applied to MCMC chains that were originally generated using uniform (“uninformative”) priors on $(w_0, w_a)$, allowing us to incorporate the theoretical model's predictions without re-running the original MCMC.

The NF is implemented with the \texttt{nflows} library~\cite{2023JOSS....8.5361S}. The base distribution is a conditional diagonal Gaussian whose parameters are predicted by a fully connected residual network. The transformation from the base to the target distribution consists of eight autoregressive coupling layers, alternating between masked affine autoregressive transforms~\cite{2017arXiv170507057P} and masked piecewise rational quadratic autoregressive transforms~\cite{2019arXiv190604032D} with eight bins and linear tails. Between successive coupling layers, a reverse permutation of $(w_0, w_a)$ is applied to enhance expressivity. Each coupling layer conditions on the three-dimensional input, employs 128 hidden units per block, and uses three residual blocks with ReLU activations. The NF is trained by minimizing the mean negative log-likelihood
\begin{equation}
    \mathcal{L} = - \frac{1}{N} \sum_{i=1}^N \log p_\theta(w_{0,i}, w_{a,i} \mid \omega_{c,i}, \omega_{b,i}, H_{0,i}),
\end{equation}
using the Adam optimizer~\cite{Kingma:2014vow} with a learning rate of $3\times 10^{-4}$ and a \texttt{ReduceLROnPlateau} scheduler. Training is performed in batches of 128 samples for up to $5000$ epochs on a NVIDIA A100 GPU, with early stopping after $100$ epochs if there is no improvement in the validation loss. All parameters are standardized prior to training, and the best-performing model checkpoint is used in subsequent analysis. In Fig.~\ref{fig:nf_loss} we present the training schedule for both of the NFs used in this work, demonstrating that our architecture does not overfit the training sample. For calculating the validation loss we withhold 20\% of the samples.
\clearpage
\section{Mapping to $(w_0,\,w_a)$ for Exponential Quintessence}

\begin{figure}[ht!]
    \centering
    \includegraphics[width=.5\linewidth]{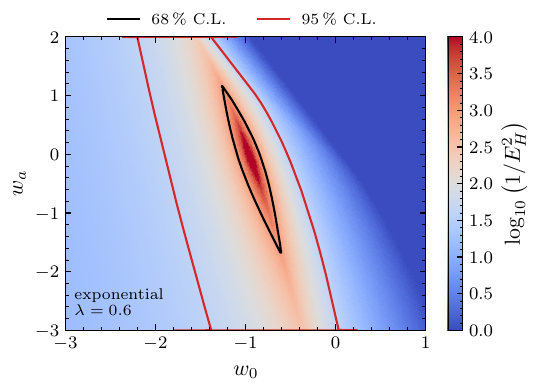}
    \caption{Same as Fig.~\ref{fig:1} for a exponential quintessence model, Eq.~\eqref{eq:exp}, with $\lambda=0.6$.}
    \label{fig:A1}
\end{figure}

\begin{figure*}[ht!]
    \centering    \includegraphics[width=0.95\linewidth]{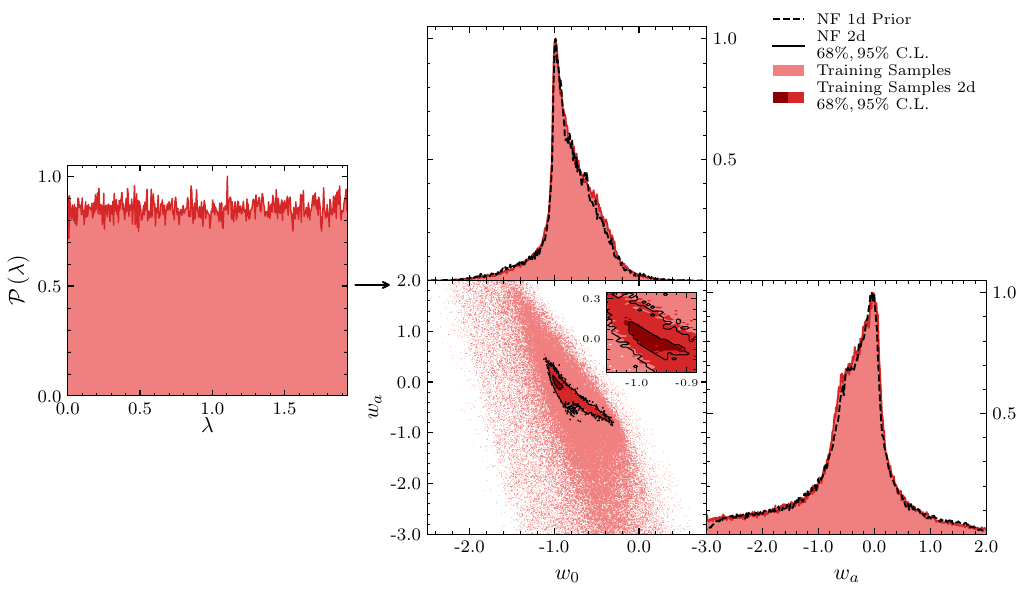}
    \caption{Same as Fig.~\ref{fig:NF_priors} for a exponential quintessence model, Eq.~\eqref{eq:exp}. The normalizing flow (black lines) accurately reproduces the training data distributions, demonstrating robust learning across both quintessence model classes. As with the hilltop case, the mapped $(w_0, w_a)$ distributions are distinctly non-uniform despite originating from flat priors on $\lambda$. However, the exponential potential produces a broader peak around $(w_0, w_a) \approx (-1, 0)$, with the distribution extending more prominently toward smaller values of $w_0$ and more negative values of $w_a$, reflecting the fundamentally different thawing dynamics of exponential quintessence compared to hilltop models. }
    \label{fig:A2}
\end{figure*}
\clearpage
\section{Comprehensive Summary of Cosmological Costraints}

\begin{table}[hb!]
	\centering
	\subfloat[Uniform Prior on $(w_0,\,w_a)$]{%
		\begingroup
		\sisetup{group-digits=false}
		\renewcommand{\arraystretch}{1.2}
		\begin{tabular}{@{}l@{\hspace{2em}}c@{\hspace{2em}}c@{\hspace{2em}}c@{\hspace{2em}}c@{}}
			\toprule
			Parameter & DESI BAO + {\it Planck} & + Pantheon+ & + Union3 & + DESY5 \\
			\midrule[0.065em]
			$100\,\omega_b$ & $2.218 \pm 0.013$ & $2.223 \pm 0.013$ & $2.220 \pm 0.013$ & $2.221 \pm 0.013$ \\
			$\omega_c$ & $0.1194 \pm 0.0009$ & $0.1187 \pm 0.0008$ & $0.1192 \pm 0.0008$ & $0.1191 \pm 0.0008$ \\
			$100\,\theta_s$ & $1.04176 \pm 0.00023$ & $1.04182 \pm 0.00023$ & $1.04179 \pm 0.00023$ & $1.04180 \pm 0.00023$ \\
			$\ln\left(10^{10}A_s\right)$ & $3.033 \pm 0.014$ & $3.038 \pm 0.014$ & $3.035 \pm 0.014$ & $3.036 \pm 0.014$ \\
			$n_{\rm s}$ & $0.9638 \pm 0.0038$ & $0.9650 \pm 0.0037$ & $0.9645 \pm 0.0037$ & $0.9646 \pm 0.0036$ \\
			$\tau$ & $0.0508 \pm 0.0071$ & $0.0536 \pm 0.0071$ & $0.0522 \pm 0.0070$ & $0.0526 \pm 0.0070$ \\
			$w_0$ & $-0.45\pm 0.21$ & $-0.84\pm 0.06$ & $-0.68\pm 0.09$ & $-0.76\pm 0.06$ \\
			$w_a$ & $-1.64\pm 0.59$ & $-0.59^{+0.21}_{-0.19}$ & $-1.03^{+0.30}_{-0.27}$ & $-0.83^{+0.23}_{-0.21}$ \\
			$H_0\,\left[{\rm km}\,{ \rm s}^{-1}\,{\rm Mpc}^{-1}\right]$ & $63.8^{+1.8}_{-2.1}$ & $67.5\pm 0.6$ & $66.0\pm 0.9$ & $66.7\pm 0.6$ \\
			$\Omega_m$ & $0.350\pm 0.022$ & $0.311\pm 0.006$ & $0.327\pm 0.009$ & $0.319\pm 0.006$ \\
			\bottomrule
		\end{tabular}
		\endgroup
	}\\[4pt]
	\subfloat[Exponential Potential Prior on $(w_0,\,w_a)$]{%
		\begingroup
		\sisetup{group-digits=false}
		\renewcommand{\arraystretch}{1.2}
		\begin{tabular}{@{}l@{\hspace{2em}}c@{\hspace{2em}}c@{\hspace{2em}}c@{\hspace{2em}}c@{}}
			\toprule
			Parameter & DESI BAO + {\it Planck} & + Pantheon+ & + Union3 & + DESY5 \\
			\midrule[0.065em]
			$100\,\omega_b$ & $2.221 \pm 0.013$ & $2.223 \pm 0.013$ & $2.222 \pm 0.013$ & $2.223 \pm 0.013$ \\
			$\omega_c$ & $0.1190 \pm 0.0009$ & $0.1186 \pm 0.0009$ & $0.1189 \pm 0.0009$ & $0.1191 \pm 0.0008$ \\
			$100\,\theta_s$ & $1.04180 \pm 0.00023$ & $1.04184 \pm 0.00024$ & $1.04181 \pm 0.00023$ & $1.04181 \pm 0.00023$ \\
			$\ln\left(10^{10}A_s\right)$ & $3.036 \pm 0.014$ & $3.040 \pm 0.014$ & $3.037 \pm 0.014$ & $3.037 \pm 0.014$ \\
			$n_{\rm s}$ & $0.9648 \pm 0.0038$ & $0.9658 \pm 0.0037$ & $0.9650 \pm 0.0037$ & $0.9651 \pm 0.0037$ \\
			$\tau$ & $0.0526 \pm 0.0072$ & $0.0550 \pm 0.0073$ & $0.0532 \pm 0.0071$ & $0.0535 \pm 0.0071$ \\
			$w_0$ & $-0.63^{+0.15}_{-0.25}$ & $-0.87\pm 0.06$ & $-0.73\pm 0.09$ & $-0.78\pm 0.06$ \\
			$w_a$ & $-1.12^{+0.68}_{-0.43}$ & $-0.46\pm 0.22$ & $-0.87\pm 0.29$ & $-0.73^{+0.25}_{-0.22}$ \\
			$H_0\,\left[{\rm km}\,{ \rm s}^{-1}\,{\rm Mpc}^{-1}\right]$ & $65.4^{+2.2}_{-1.6}$ & $67.5\pm 0.6$ & $66.2\pm 0.8$ & $66.8\pm 0.6$ \\
			$\Omega_m$ & $0.333^{+0.016}_{-0.024}$ & $0.311\pm 0.006$ & $0.323\pm 0.009$ & $0.318\pm 0.006$ \\
			\bottomrule
		\end{tabular}
		\endgroup
	}\\[4pt]
	\subfloat[Hilltop Potential Prior on $(w_0,\,w_a)$]{%
		\begingroup
		\sisetup{group-digits=false}
		\renewcommand{\arraystretch}{1.2}
		\begin{tabular}{@{}l@{\hspace{2em}}c@{\hspace{2em}}c@{\hspace{2em}}c@{\hspace{2em}}c@{}}
			\toprule
			Parameter & DESI BAO + {\it Planck} & + Pantheon+ & + Union3 & + DESY5 \\
			\midrule[0.065em]
			$100\,\omega_b$ & $2.222 \pm 0.013$ & $2.224 \pm 0.013$ & $2.221 \pm 0.014$ & $2.222 \pm 0.013$ \\
			$\omega_c$ & $0.1189 \pm 0.0009$ & $0.1187 \pm 0.0009$ & $0.1190 \pm 0.0009$ & $0.1190 \pm 0.0008$ \\
			$100\,\theta_s$ & $1.04182 \pm 0.00023$ & $1.04184 \pm 0.00024$ & $1.04181 \pm 0.00023$ & $1.04180 \pm 0.00023$ \\
			$\ln\left(10^{10}A_s\right)$ & $3.037 \pm 0.014$ & $3.040 \pm 0.014$ & $3.036 \pm 0.014$ & $3.036 \pm 0.014$ \\
			$n_{\rm s}$ & $0.9655 \pm 0.0037$ & $0.9650 \pm 0.0037$ & $0.9648 \pm 0.0038$ & $0.9647 \pm 0.0036$ \\
			$\tau$ & $0.0533 \pm 0.0071$ & $0.0548 \pm 0.0072$ & $0.0529 \pm 0.0072$ & $0.0529 \pm 0.0070$ \\
			$w_0$ & $-0.78^{+0.09}_{-0.22}$ & $-0.89\pm 0.06$ & $-0.76^{+0.09}_{-0.10}$ & $-0.79\pm 0.06$ \\
			$w_a$ & $-0.74^{+0.61}_{-0.30}$ & $-0.43\pm 0.22$ & $-0.80\pm 0.32$ & $-0.72^{+0.25}_{-0.21}$ \\
			$H_0\,\left[{\rm km}\,{ \rm s}^{-1}\,{\rm Mpc}^{-1}\right]$ & $66.9^{+2.1}_{-1.0}$ & $67.8^{+0.6}_{-0.5}$ & $66.7\pm 0.9$ & $67.0\pm 0.6$ \\
			$\Omega_m$ & $0.318^{+ 0.009}_{-0.021}$ & $0.308\pm 0.005$ & $0.320\pm 0.009$ & $0.316\pm 0.006$ \\
			\bottomrule
		\end{tabular}
		\endgroup
	}
	\caption{Summary table of all cosmological parameter constraints from DESI DR2 BAO in combination with {\it Planck} and additional supernova datasets, under different priors, in the $w_0 w_a$CDM model. The table includes all varied parameters during the MCMC analyses as well as key derived parameters. Results quoted for all parameters are the marginalized posterior means and their corresponding 68\% credible intervals.}
	\label{tab:analysis_all}
\end{table}
\end{document}